\documentclass[12pt,epsf]{article}
\usepackage{epsfig}

\newcommand{\onefigure}[2]{\begin{figure}[htbp]
\begin{center}\leavevmode\epsfbox{#1.eps}\end{center}\caption{#2\label{#1}}
\end{figure}}

\renewcommand{\thanks}[1]{\footnote{#1}} 

\newcommand{\be}{\begin{equation}}
\newcommand{\ee}{\end{equation}}
\newcommand{\bea}{\begin{eqnarray}}
\newcommand{\eea}{\end{eqnarray}}

\begin{document}

\begin{flushright}
SLAC-PUB-11437\\
19 August 2005\\
\end{flushright}

\bigskip\bigskip
\begin{center}
{\bf\large Evidence for a Cosmological Phase Transition
on the TeV Scale\footnote{\baselineskip=12pt Work supported by Department
of Energy contract DE--AC03--76SF00515.}}
\end{center}

\begin{center}
James V. Lindesay\footnote{Permanent address, Physics Department, Howard University,
Washington, D.C. 20059}, jlslac@slac.stanford.edu\\
H. Pierre Noyes, noyes@slac.stanford.edu \\
Stanford Linear Accelerator Center MS 81,
Stanford University \\
2575 Sand Hill Road, Menlo Park CA 94025\\
\end{center}

\begin{center}
{\bf Abstract}
\end{center}
Examining the reverse evolution of the
universe from the present, long before reaching Planck density
dynamics one expects major modifications from the de-coherent
thermal equations of state, suggesting a prior phase that
has macroscopic coherence properties.
The assumption that the phase transition occurs during the radiation
dominated epoch, and that zero-point motions drive the
fluctuations associated with this transition, specifies
a class of cosmological
models in which the cosmic microwave background fluctuation
amplitude at last scattering is approximately $10^{-5}$.
Quantum measurability constraints (eg. uncertainly relations)
define cosmological scales
whose expansion rates can be at most luminal.
Examination of these constraints for the observed dark energy density
establishes a time interval from the transition to the present.
It is shown that the dark energy can consistently be interpreted
as due to the vacuum energy of collective gravitational modes
which manifest as the zero-point motions of coherent Planck
scale mass units prior to the gravitational quantum de-coherence
of the cosmology.
A scenario is suggested that connects microscopic physics to the
relevant cosmological scale.

\bigskip

\setcounter{equation}{0}
\section{Introduction \label{intro}}
\indent

There is general (although not universal) agreement among physical
cosmologists that the current expansion phase in the evolution of
our universe can be extrapolated back toward an initial state of
compression so extreme that we can neither have direct laboratory
nor indirect (astronomical) observational evidence for the laws of
physics needed to continue that extrapolation.  However, much
can be ascertained about the physics that is transmitted through
the expansion by the evidence available to modern astrophysical
measurements.  For an excellent review of the status of
observational cosmology, see reference \cite{Paddy}.

Here, it is assumed that the experimental evidence for currently accepted
theories of particle physics is relevant in the TeV range.  It is
further assumed that the current understanding of general
relativity as a gravitational theory is adequate over the same
range, and consequently that the cosmological Friedman-Lemaitre
(FL) dynamical equations are reliable guides once 
the observational regime has been reached where the homogeneity and
isotropy assumptions on which those equations are based become
consistent with astronomical data to requisite accuracy. 
The elementary particle theories usually employed in relativistic
quantum field theories have well defined transformation properties
in the flat Minkowski space of special relativity; 
their fundamental principles are assumed to apply on coordinate backgrounds
with cosmological curvature. There is direct experimental
evidence that quantum mechanics does apply in the background space
provided by the Schwarzschild metric of the Earth from experiments by Overhauser and
collaborators\cite{Over74,Over75}. These experiments show that coherent
self-interference of single neutrons changes as expected
when the plane of the two interfering paths is rotated from being
parallel to being perpendicular to the ``gravitational field" of
the Earth.  These were the first measurements requiring both Newton's constant
and Planck's constant.  They provide a verification of the principle of equivalence
for quantum systems. 

It is therefore expected
that during some period in the past, quantum coherence of
gravitating systems should have qualitatively altered the thermodynamics
of the cosmology.  Often, the onset of the importance of quantum
effects in gravitation is taken to be at the Planck scale.
However,as is the case with Fermi degenerate stars, this need not
be true of the cosmology as a whole.  Quantum coherence
refers to the entangled nature of quantum states for space-like separations.
This is made evident by superluminal correlations (without the
exchange of signals) in the observable behavior of such quantum states.
Note that the exhibition of quantum coherent behavior for gravitating systems does
\emph{not} require the quantization of the gravitation field.

The (luminal) horizon problem for present day cosmology
arises from the large
scale homogeneity and isotropy of the observed universe.  Examining the
ratio of the present conformal time $\eta_o$ (which multiplied by c is the distance a photon can travel in a given time)
with that during recombination $\eta_* , {\eta_o \over \eta_*} \sim 100$,
the subsequent expansion is expected to imply that light from the cosmic microwave background would
come from $100^3=10^6$ luminally disconnected regions. 
Yet, angular correlations of the fluctuations across the whole sky have been
accurately measured by several experiments\cite{WMAP}.  The observation
of these correlations provides evidence for
a space-like coherent phase associated with the
cosmological fluctuations that produced the Cosmic Microwave Background (CMB).

The approach taken here will not depend on the present particle horizon scale,
which is an accident of
history.  It will be argued that the equilibration of microscopic interactions
can only occur on cosmological scales consistent with quantum
measurement constraints\cite{ANPA04}.
Global quantum coherence on larger scales solves the horizon problem,
since quantum correlations are in this sense supraluminal.

The luminosities of distant Type Ia
supernovae show that the rate of expansion of the
universe has been accelerating for several
giga-years\cite{TypeIa}. This conclusion is independently
confirmed by analysis of the Cosmic Microwave Background (CMB)
radiation\cite{PDG,WMAP}.  Both results are in quantitative
agreement with a (positive) cosmological constant fit to the data.
The existence of a cosmological constant / dark energy density
defines a length scale
that should be consistent with those scales generated by the
microscopic physics, and must be incorporated in any
description of the evolution of the universe.   When the dynamics
of the cosmology is made consistent with this scale, it is expected
that energy scales associated with
the usual microscopic interactions of relativistic quantum
mechanics (QED, QCD, etc) cannot contribute to cosmological
(gravitational) affects when the relevant Friedmann-Robertson-Walker
(FRW) scale expansion rate is supra-luminal\cite{Rdot=c}, $\dot{R}_\epsilon>c$.  
Quantum measurement arguments will be made below for quantized
energy units $\epsilon$ requiring that the scales associated with those
energies, $R_\epsilon$, must satisfy $\dot{R}_\epsilon \leq c$.
The gravitationally coherent
cosmological dark energy density should decouple from the thermalized
energy density in the Friedmann-Lemaitre(FL) equations when the
relevant FRW scale expansion rate is no longer supra-luminal. 
A key assumption in this paper is that this decoupling corresponds
to a phase transition from some form of a macroscopic coherent state
to a state understood in terms of late time observations.
The form of the FL equations \ref{Hubble_eqn} and \ref{acceleration}
is clearly scale invariant in $R$ (other than the term involving
spatial curvature) for given intensive densities, which might
lead one to question the physical meaning of $R$.  However, this
scale invariance does \emph{not} negate an assignment of a meaningful
FRW scale parameter whose dynamics is governed by this equation.
The scale below which the cosmology is no longer homogeneous
is an example of a well defined particular scale of cosmological
significance whose evolution is governed by the interplay of the
dynamics of the FL equations with the
density perturbations.  Likewise, any infrared
cutoff wavelength associated with the physics that generates those
perturbations is an example of such a scale.

The basic assumption in this paper is that the
quantum zero-point energies whose effects on the subsequent
cosmology are fixed by the phase transition
mentioned in the previous paragraph
are to be identified with the cosmological constant, or ``dark energy".
It has been shown elsewhere\cite{JLHPNEJ}
that the expected amplitude of fluctuations driven by
dark energy as the energy density de-coheres
is of the order needed to evolve
into the fluctuations observed in CMB
radiation and in galactic clustering. From the time of this phase
transition until this scale factor expansion rate again becomes
supra-luminal due to the cosmological
constant, the usual expansion rate evolution predicted by FL
dynamics, including a cosmological constant, is expected to hold.

Since this paper will eventually identify ``dark energy" as a
particular ``vacuum energy" driven by zero-point motions, it is
worthwhile to examine the physics behind other systems that manifest
vacuum energy.  One physical system in which vacuum energy density directly manifests is the Casimir effect\cite{Casimir}.
Casimir considered the change in the vacuum energy due to the placement of
two parallel plates separated by a distance $a$.  He calculated an energy per unit area of the form
\be
{ {1 \over 2} \left ( \sum_{modes} \hbar c k_{plates} -
\sum_{modes} \hbar c k_{vacuum}  \right ) \over A} \: = \:
- {\pi^2 \over 720} {\hbar c \over a^3}
\label{Casimir}
\ee
resulting in an attractive force of given by
\be
{F \over A}  \: = \:
- {\pi^2 \over 240} {\hbar c \over a^4}
\: \cong \: {- 0.013 \, dynes \over (a / micron)^4  } cm^{-2},
\ee
independent of the charges of the sources.  Although
the effect does not depend on the electromagnetic
coupling strength of the sources, it does depend on
the nature of the interaction and configuration. 
Lifshitz and his collaborators\cite{Lifshitz} demonstrated that the Casimir force
can be thought of as the superposition of the van der Waals attractions
between individual molecules that make up the attracting media.  This allows
the Casimir effect to be interpreted in terms of the zero-point motions of the
sources as an alternative to vacuum energy. 
At zero temperature, the coherent zero-point motions of
source currents on opposing plates correlate in a manner resulting
in a net attraction, whereas if the motions were independently random,
there would be no net attraction.  On dimensional grounds, one can
determine that the number of particles per unit area undergoing
zero-point motions that contribute to the Casimir result vary as $a^{-2}$.
Boyer\cite{Boyer} and others subsequently
derived a repulsive force for a spherical geometry of the form
\be
{1 \over 2} \left ( \sum_{modes} \hbar c k_{sphere} -
\sum_{modes} \hbar c k_{vacuum}  \right )  \: = \:
{ 0.92353 \hbar c \over a   }.
\ee
This shows that the change in electromagnetic vacuum energy \textit{is}
dependent upon the geometry of the boundary conditions.  Both
predictions have been confirmed experimentally.  It is important to note
that this energy grows inversely with the geometrical scale
$E_{Casimir} \sim {\hbar c \over a}$.

Generally, correlated zero-point motions can be used to describe
vacuum energy effects in the Casimir effect.  As expressed by
Daniel Kleppner\cite{Kleppner},
\begin{quotation}
The van der Waals interaction is generally described in terms of
a correlation between the instantaneous dipoles of two atoms
or molecules.  However, it is evident that one can just as easily
portray it as the result of a change in vacuum energy due to
an alteration in the mode structure of the system.  The two
descriptions, though they appear to have nothing in common,
are both correct.
\end{quotation}
Additionally, as pointed out by Wheeler and Feynmann\cite{Wheeler},
and others\cite{LN2},
one cannot unambiguously separate the properties of
fields from the interaction of those fields with their sources
and sinks.
Since there are no manifest boundaries in the description of
early cosmology presented here, it is much more convenient
to examine the effects of any gravitational ``vacuum energy'' in terms of
the correlated zero-point motions of the sources of those
gravitational fields.  In what follows, the zero-point motions of
coherent sources will be considered to correspond to the
vacuum energies of the associated quanta. 
This point is elaborated at the start of Sec. \ref{PTfluc}

The introduction of energy density $\rho$ into Einstein's equation introduces
a preferred rest frame with respect to any normal energy density. 
However, as can be seen in the Casimir effect,
the vacuum need not exhibit velocity dependent effects which would break
Lorentz invariance.  
Although a single moving mirror has been predicted to
experience dissipative effects from the vacuum due to its motion,
these effects can be shown to be of 5th order in time derivatives
of position\cite{Genet}.

Another system which manifests physically measurable effects due to zero-point energy is liquid $^4$He.
One sees that this is the case by noting that atomic radii are related to atomic volume $V_a$
(which can be measured) by $R_a \sim V_a ^{1/3}$.  The uncertainty relation gives momenta of the order
$\Delta p \sim \hbar / V_a^{1/3}$.  Since the system is non-relativistic, one estimates the zero-point kinetic
energy to be of the order $E_o \sim {(\Delta p)^2 \over 2 m_{He}} \sim {\hbar ^2 \over 2 m_{He} V_a ^{2/3}}$.
The minimum in the potential energy is located around $R_a$, and because of the low mass of $^4$He, the
value of the small attractive potential is comparable to the zero-point kinetic energy.  Therefore, this bosonic
system forms a low density liquid.  The lattice spacing for solid helium would be expected to be even
smaller than the average spacing for the liquid.  This means that a large external pressure is necessary to
overcome the zero-point energy in order to form solid helium.  Generally
zero-point motions correspond to the saturation of the triangle inequality as
expressed in the uncertainty principle.

Applying this reasoning to relativistic gravitating mass units with quantum coherence within the
volume generated by a Compton wavelength $\lambda_m ^3$, the zero point momentum is expected
to be of order $p \sim {\hbar \over V^{1/3}} \sim {\hbar \over \lambda_m}$.  This gives a zero point
energy of order $E_0 \approx \sqrt{2} m c^2$.  If we estimate a mean field potential from the
Newtonian form $V \sim -{G_N m^2 \over \lambda_m}=-{m^2 \over M_P ^2}mc^2<<E_0$, it is evident that
the zero point energy will dominate the energy of such a system.

As a further illustration of the expectation of the manifestation of
macroscopic quantum effects on a cosmological scale, consider
particulate dark matter.  If the dark matter behaves as a bosonic
particle, one can estimate the critical number density by
examining a free bose gas.  For non-relativistic dark matter,
this critical density is reached when thermal modes can no
longer accomodate a distribution of all of the particles, forcing
macroscopic occupation of the lowest energy state.  
For particles of mass $m$, this density satisfies\cite{standardtext}
\be
{N_{DM} \over V} = {\zeta(3/2) \Gamma(3/2) \over (2 \pi)^2 \hbar^3}
(2 m k_B T_{crit})^{3/2} \quad , \quad
\rho_{DM} \cong {N_{DM} \over V} m c^2 .
\ee
This equation can be used to determine the relationship between
the critical temperature and the redshift associated
with Bose condensation of such particles, given by
\be
k_B T_{crit}= {1 \over 2} {\rho_{DM} ^{2/3} \over (m c^2)^{5/3}}
{(2 \pi)^2 (\hbar c)^3 \over \zeta(3/2) \Gamma(3/2)} \approx 
1.7 \times 10^{-31} GeV {z^2 \over g_m ^{2/3}}
\left ( {GeV \over m c^2}  \right )^{5/3}.
\ee
The temperature of the photon gas is expected to scale with
the redshift when the appropriate pair creation threshold
effects are properly incorporated.  Setting the critical
temperature the same as the photon temperature gives
an estimate of the onset of cosmological bose condensation
for such particles, given by
\be
z_{crit} \cong 1.4 \times 10^{18} 
\left( {g(T_{\gamma o}) \over g(T_{crit})}  \right ) ^{1/3}
g_m ^{2/3} \left ( {m c^2 \over GeV}  \right )^{5/3},
\ee
where $g(z)$ counts the number of low mass degrees of
freedom available at redshift $z$.  
The transition occurs
at non-relativistic energies as long as the particulate mass
satisfies $m>{15eV \over g_m}$.  
Some authors have suggested
that present day Bose condensation addresses some of the
problems associated with the dark matter\cite{Silverman}.

Adiabatic expansion is expected to preserve the ratio of
particulate dark matter particle number to photon number.
Relating photon energy density to number density gives
\be
{N_{DM} \over N_\gamma} = 
{ \zeta(4) \Gamma(4) \over  \zeta(3) \Gamma(3)}
\left ( {\Omega_{DMo} \over \Omega_{\gamma o}} \right )
{k_B T_{\gamma o} \over m c^2} \cong 
3.7 \times 10^{-9} \left( {GeV \over m_{DM} c^2}  \right ) .
\ee
This represents a phenomenological handle on any
cosmological affects due to thermalization of
particulate bose condensed dark matter.

The Friedman-Robertson-Walker (FRW) metric for a homogeneous
isotropic cosmology is given by
\be
ds^2 \: = \: c^2 dt^2 \, + \, R^2 (t) \left (
{dr^2 \over 1-k \, r^2} + r^2 d \theta ^2
+ r^2 sin^2 \theta \, d \phi ^2
\right )
\label{FRWmetric}
\ee
In a radiation-dominated universe this backward extrapolation in time
(which taken literally {\it must} terminate when \emph{any}
FRW scale factor $R(t)$ goes to zero and its
time rate of change $\dot R(t)$ goes to infinity) is guaranteed to
reach the speed of light  $\dot R(t_c)=c$  at some finite time
$t_c$ when the scale factor $R(t_c)$ still has a small, but
finite, value.  Here the FRW scale factor $R$ is taken to have dimensions of length
(not to be the dimensionless scale relative to the present horizon
$a(t) \equiv R(t)/R_o$), with $dr$ being dimensionless.
One aim of this paper is to show that for quantized energies, a microscopically
motivated cosmological scale can be defined which indicates a macroscopic
phase transition after which that scale expands at a sub-luminal rate.

The general approach used here has been to start from well
understood macrophysics,
assume that the physics of a cosmological phase transition defines an
FRW scale parameter, and examine
cosmological physics at the
time when the physical consistency of the thermal state
of the cosmology is called into question. For times after that transition
there is general confidence that well
understood micro- \emph{and} macro-physics are valid
at the cosmological level.
The relevant FRW scale parameter must be expressed in terms of the
scales of microscopic physics. The calculations presented here will not
use the present horizon (Hubble) scale other than to evaluate
observed phenomenological parameters.

The ``horizon problem"
addresses the extreme uniformity of the Cosmic Microwave Background
Radiation across multitudes of space-like separated regions, and the
space-like correlation in the phase of the fluctuation across the entire
sky, despite these regions being luminally disconnected at the time
of last scattering, when these photons were produced. 
In quantum physics, such phenomena are not unusual.
Using the usual vacuum state in Minkowski space-time, the equal time correlation function 
${<vac|\Psi(x,y,z,t) \, \Psi(x',y',z',t)|vac>}$ of a quantum field $\Psi(\vec{x})$
does not vanish for space-like separations.  For example, for massless scalar fields, 
$<vac|\Psi(\vec{x}) \, \Psi(\vec{y})+\Psi(\vec{y}) \, \Psi(\vec{x})|vac>={1 \over 4 \pi^2 s^2}$, where
the proper distance satisfies $s^2 = |\mathbf{x}-\mathbf{y}|^2 -
(x^0-y^0)^2$, which falls off with the inverse square of the distance between the points. 
Since the vacuum expectation value of the field $\Psi$ vanishes in the usual case, this
clearly requires space-like correlations, ie
\be
\begin{array}{c}
{1 \over 4 \pi ^2 s^2} =
<vac|\Psi(\vec{x}) \, \Psi(\vec{y})+\Psi(\vec{y}) \, \Psi(\vec{x})|vac>\not= \\
2 <vac|\Psi(\vec{x}) |vac><vac| \Psi(\vec{y})|vac> = 0.
\end{array}
\ee
However, since the commutator of the field \emph{does} vanish for
space-like separations, a measurement at $\vec{y}$ cannot change
the probability distribution at $\vec{x}$.
In the approach here taken, global gravitational
coherence solves (or defers) the
horizon problem because the gravitational correlations implied by
the FL equations \emph{are} space-like; it is hypothesized that the same
will be true of any type of dark energy of gravitational origin
to be considered. 

\setcounter{equation}{0}
\section{Estimate of Density Fluctuations
\label{PTfluc}}
\indent

As has been argued, vacuum energy can often be thought of as resulting from
the zero-point motions of the sources\cite{Lifshitz}, and this is expected
to be true of gravitational interactions. 
This argument is supported by the calculation of Bohr
and Rosenfeld\cite{BohrRosenfeld}, who minimized the effect of a 
classical measurment of an electric field averaged over a finite volume
on the value of a magnetic field at right angles averaged over (i) a
non-overlapping volume, and (ii) an overlapping volume, and {\it vice versa}. 
When this minimum disturbance is put equal to the minimum uncertainty
which the uncertainty principle allows for the measurement they
reproduce the result of averaging the quantum mechanical commutation
relations over the corresponding volumes.  Such arguments can be extended
to the corresponding case when the sources and detectors are
gravitational.

As the zero-point motions de-cohere and become localized,
the deviations from uniformity are expected to appear as
fluctuations in the cosmological energy density. 
Since these motions are inherently
a quantum effect, one expects the fluctuations to exhibit the
space-like correlations consistent with a quantum phenomenon. 
Measurable effects of quantum mechanical de-coherence
are expected to manifest stochastically.

As an intuitive guide into how dark energy might freeze
out as as system de-coheres, consider a uniform distribution
of non-relativistic masses $M$ interacting pairwise through simple harmonic
potentials at zero temperature.  
If charges were used as field sources rather than masses,
as explained above the
uncertainty relations associated with the positions and momenta
of the charges result in averaged commutation relations between
the electric and magnetic fields which are classically produced by
those charges. 
Zero-point motions of the oscillators
give equal partitioning of energies ${1 \over 4} \hbar \omega$ to the
kinetic and potential components.  The 
correlations of motions of the masses do not vanish, although time
averages do vanish.  If the masses evaporate, the
kinetic component is expected to drive density fluctuations of
an expanding gas, whereas
the potential components remain in the springs (as zero point
tensions) of fixed density, as illustrated in Figure \ref{DCall}.
\onefigure{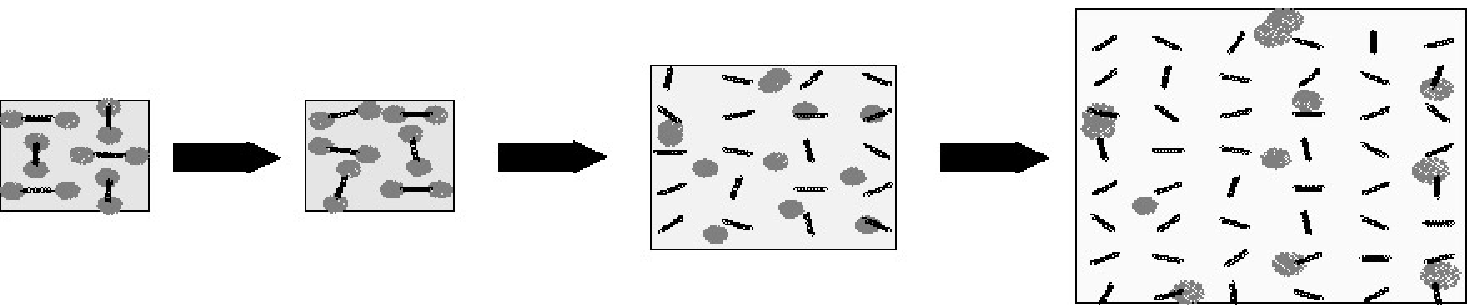}{Early quantum stage undergoing zero-point
fluctuations during de-coherence, with evaporation and
subsequent expansion of thermal energy density associated
with masses which were previously coherently attached
via the springs.  Potential
energy density gets frozen in during evaporation, whereas 
kinetic energy density drives density fluctuations.}
The compressional energy is expected to be frozen in as
dark energy in the background during the phase transition
on scales larger than the de-coherence scale. 
The mass units which were undergoing zero-point motions
no longer behave as coherent masses after evaporation,
and only reflect their former state through the density
inhomogeneities generated during the evaporation process,
and no longer couple to the frozen-in potential component of
the zero point energy.  Such an interpretation would clearly
represent the source of a cosmological constant as a frozen energy density,
\emph{not} a fixed background metric variation.

An interacting sea of the quantum fluctuations due to
zero point motions should exhibit local statistical variations in
the energy.  For a sufficiently well defined phase state,
one should be able to use counting
arguments to quantify these variations\cite{JLHPNEJ, Poster}.
The dark energy $E_\Lambda$ is expected to have uniform density,
and to drive fluctuations of the order
\be
<(\delta E)^2> \: = \: E_\Lambda ^2 {d \over dE_\Lambda} <E>
\ee
Given an equation of state $<E> \sim (E_\Lambda )^b$,
the expected fluctuations satisfy
\be
{<(\delta E)^2> \over <E>^2} 
\: = \: b \, {E_\Lambda \over <E>} \, .
\label{delE}
\ee
In terms of the densities, one can directly write
${<(\delta E)^2> \over <E>^2} \: = \: {<(\delta \rho)^2> \over \rho^2}
\: = \: b \,  {\rho_\Lambda \over \rho}$.
At the time of the formation of the fluctuations, this means that the
amplitude $\delta \rho / \rho$
is expected to be of the order 
\be
\delta_{PT} \: = \: \left (  b \,{\rho_\Lambda \over \rho_{PT} } \right ) ^{1/2}  
\: = \: \sqrt{b} {R_{PT} \over R_\Lambda}
\label{DelPT}
\ee 
where $\rho_{PT}$ is the cosmological energy density
at the time of the phase transition that decouples the dark energy
and $\Lambda = 8 \pi G_N \rho_\Lambda /c^4 =3/R_\Lambda ^2$
is the cosmological constant.

For adiabatic perturbations (those that fractionally perturb the number
densities of photons and matter equally),
the energy density fluctuations grow according to\cite{PDG}
\be
\delta \: = \: \left \{
\begin{array}{cc}
\delta_{PT}  \left ( {R(t) \over R_{PT}} \right ) ^2  &
radiation-dominated \\
\delta_{eq} \left ( {R(t) \over R_{eq}} \right ) &
matter-dominated
\end{array}
\right . ,
\ee
which gives an estimate for the scale of fluctuations
at last scattering from fluctuations during the
phase transition expressed by
\be
\begin{array}{l}
\delta_{LS} \, = \, { (1+z_{PT}) ^2 \sqrt{b} \over (1+z_{eq}) (1+ z_{LS}) }
\left (  {\rho_\Lambda \over \rho_{PT} } \right ) ^{1/2} \\ \\  \quad \quad \cong
{1 \over 1 + z_{LS}} \sqrt{{b \, \Omega_{\Lambda o}
\over (1-\Omega_{\Lambda o}) (1+z_{eq})}} \cong 2.5 \times 10^{-5} \sqrt{b}, 
\end{array}
\ee 
where a spatially flat cosmology and radiation domination has been assumed.
This estimate is independent of the density during
the phase transition $\rho_{PT}$, and is of the order observed for
the fluctuations in the CMB (see \cite{PDG} section 23.2 page 221).

A tentative conclusion is
that any \emph{phenomenological} theory that accepts
a constant dark
energy density $\rho_\Lambda={\Lambda \over 8 \pi G_N}$ 
as a reasonable way to fit the
observational data indicating a flat, accelerating universe
over the relevant range of red shifts,
and which also  assumes some sort of phase transition that
decouples the residual dark energy from the subsequent
dynamics of the energy density, will fall within a class
of theories all of which fit the magnitude of the observed
fluctuations. 

\setcounter{equation}{0}
\section{Zero-Point Motions of Quantized Energy Scales}

\subsection{Gravitating quantum energies}
\indent

For gravitating thermal systems, typical thermal energies
$k_B T_{crit}$ are given by kinetic energies for constituent
particles of mass m, which
define a thermal distance scale $R_{thermal} \approx
{\hbar c \over k_B T_{crit}}$ that satisfies
\be
R_{thermal} \lambda_m \sim (\Delta x)^2
\label{thermalscale}
\ee
in terms of the Compton wavelength of the mass and
the scale of zero-point motions of those masses.
This relationship just follows from the momentum-space
uncertainty principle.  For example, for a degenerate free
Fermi gas, the number density relationship
$n={g_m \over 6 \pi^2} (2 m \epsilon_{thermal})^{3/2}$
implies $R_\epsilon \lambda_m = {1 \over 2}
\left ( {6 \pi^2 \over g_m} \right )^{2/3} (\Delta x)^2$\cite{standardtext}.

If $\lambda_E$ represents the microscopic coherence length
of a correlated region of energy $E$, one expects a phase transition
for densities of the order 
$\rho_{PT} \: \sim \: {E \over \lambda_E ^3}$.
If densities in a region of gravitational coherence of scale
$R$ exceeds this value, one expects that
regions of quantum coherence of interaction energies of the order
of $E$ and scale $\lambda_E$ will overlap sufficiently over the
scale $R_\epsilon$ such that there will be a macroscopic quantum
system on a cosmological scale. 
\onefigure{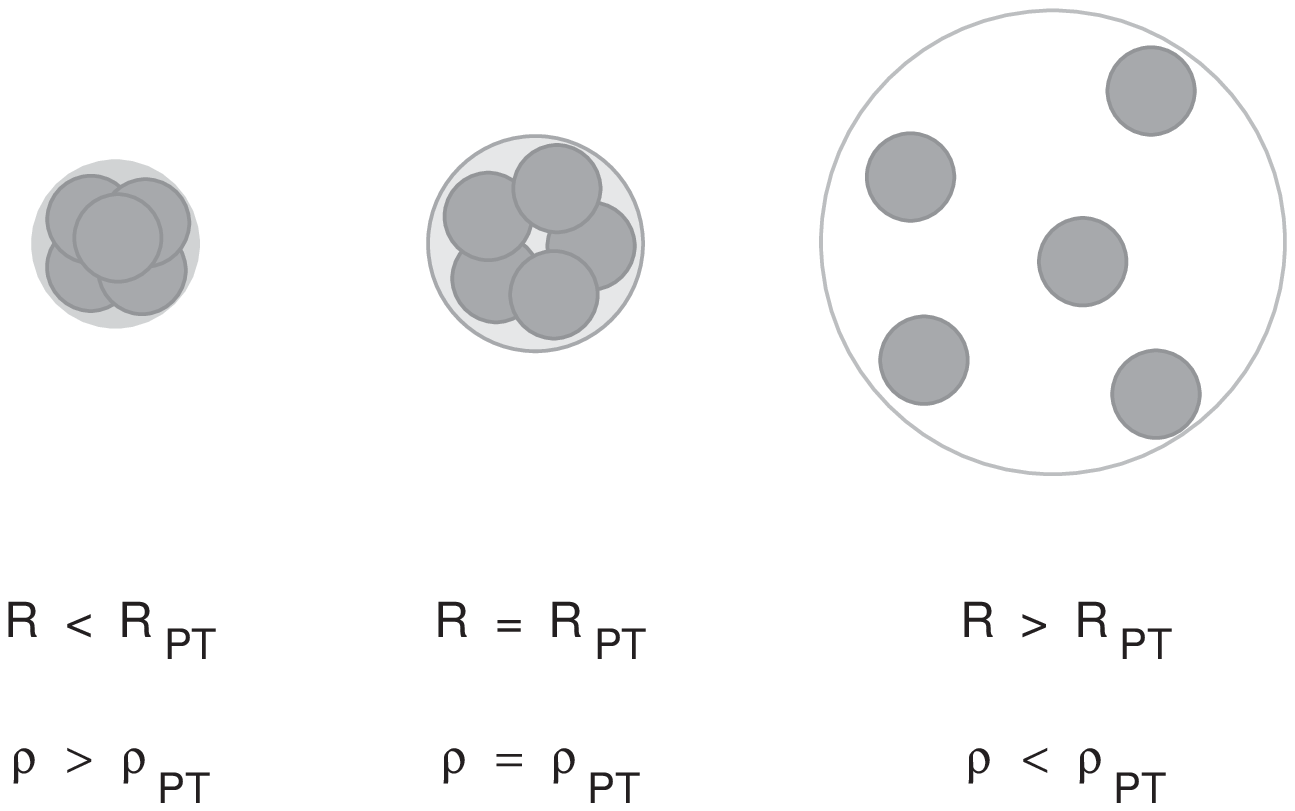}{Overlapping
regions of coherence during expansion} 
As long as the region of
gravitational coherence is of FRW scale $R \leq R_{PT}$,
cosmological manifestations of this energy are of the scale $R_{PT}$.
However, when the density of FL energy becomes less than $\rho_{PT}$,
the coherence length of those microscopic energy units given by
$\lambda_E$ is insufficient to cover the cosmological
scale, and the FL energy density will break into domains of cluster
decomposed (see ref. \cite{AKLN} for scattering theoretical
considerations) regions of local
quantum coherence. This phase transition will decouple quantum
coherence of gravitational interactions on the cosmological scale
$R_\epsilon$. At this stage (de-coherence), the cosmological
(dark) vacuum energy density $\rho_\Lambda$ associated with subtle
gravitational coherence amongst the constituent energies is frozen at the scale
determined by $R_\epsilon$. 
The cosmological dark energy
contribution to the expansion rate 
at that time is so small, and its coupling
to de-coherent FL energy so insignificant, that its value is
frozen at the value during de-coherence given by 
\be
\rho_\Lambda \: \sim \: {\epsilon \over R_\epsilon ^3} \: = \:
{\epsilon ^4 \over (\hbar c)^3}
\ee

\setcounter{equation}{0}
\section{Gravitational De-coherence From Dark Energy
\label{decohere2}}

\subsection{Quantum Measurability Constraints on Scale Expansion}
\indent

Substitution of the isotropic, homogeneous FRW metric given by
Eq. \ref{FRWmetric} into the Einstein Field equation driven by
an ideal fluid result in the Friedmann-Lemaitre(FL) equations.
The FL equations,
which relate the rate and acceleration of the expansion to the 
fluid densities, are given by
\be
H^2 (t) \: = \: \left ( {\dot{R} \over R} \right ) ^2 \: = \: {8 \pi G_N \over 3 c^2}
\left ( \rho + \rho_\Lambda  \right ) \, - \, {k c^2 \over R^2}
\label{Hubble_eqn}
\ee
\be
{\ddot{R} \over R} \: = \:
-{4 \pi G_N \over 3 c^2} (\rho + 3 P - 2 \rho_\Lambda),
\label{acceleration}
\ee
where $H(t)$ is the Hubble expansion rate, the dark energy density is given by
$\rho_\Lambda \: = \: {\Lambda c^4 \over 8 \pi G_N}$,
$\rho$ represents the FL matter-energy density,
and $P$ is the pressure.
The term which involves the spatial curvature $k$ has
explicit scale dependence on the FRW parameter $R$.
The dark energy density is assumed to make a negligible contribution to
the FL expansion during de-coherence, but will become significant as the FL energy density
decreases due to the expansion of the universe.


If $R$ is an arbitrary scale in the
Friedmann-Lemaitre equations for a spatially flat space in the
radiation-dominated epoch and a phase transition occurs at scale
$R_{PT}$ with expansion rate ${\dot R}_{PT}$,
then Eq. \ref{Hubble_eqn} gives
\be
\left ( {{\dot R}\over R} \right )^2 = {8 \pi G_N \over 3 c^4}
\rho_{PT}{R_{PT}^4\over R^4} = {{\dot R}_{PT}^2R_{PT}^2\over
R^4}
\ee
which simplifies to the form
\be
{\dot R}R = {\dot R}_{PT}R_{PT}=const.
\ee
This form can be integrated to obtain
\be
t = {R \over2 {\dot R} }
\label{timePT}
\ee
where the conditions assume that $t=0$ when $R=0$, which
smoothly defines the dynamics during this period.  Since
the approach used here involves a reverse extrapolation
within the regime of known physics into a domain where
that physics undergoes some type of phase transition or
modification, the use of the analytic behavior of parameters in the
radiation dominated regime is justified.

Any \emph{quantized} energy scale $\epsilon$ defines a
length scale $R_{\epsilon}$ by the relation
\be
\epsilon = {\hbar c \over R_{\epsilon}}. 
\ee
Using a quantized energy
scale  of order $\epsilon$ and cosmological time associated with this
scale via Eq. \ref{timePT} in the
energy-time uncertainty relation defines a constraint on
the expansion rate associated with that scale:
\be
\Delta E \Delta t \geq {\hbar \over 2}  \Rightarrow 
{\hbar c\over R_{\epsilon}} {R_\epsilon \over 2 \dot{R}_\epsilon} \geq {\hbar \over 2}
\Rightarrow \dot{R_\epsilon} \le c.
\ee 
Therefore, assuming that the scale factor at the time
of the phase transition is defined by a quantized microscopic
energy scale $R_{PT}=R_{\epsilon}$,
this scale must satisfy\cite{SLACTalk}
\be {\dot R}_{PT} \: = \: c . \ee
That is, if it is required that
ordinary, micoroscopic, event-decoherent \emph{quantum}
measurements satisfying the uncertainty principle are to be
possible \emph{after} the phase transition, $then$ the rate of
expansion of the FL scale factor associated with that phase transition
must be sub-luminal. 
If the expansion rate associated with energy $\epsilon$
is supra-luminal $\dot{R_\epsilon}>c$,
scattering states of this energy cannot form decomposed (de-coherent)
clusters of the type described in references \cite{AKLN,LMNP}
on cosmological scales, i.e. incoherent decomposed
clusters (scattering states) cannot be cosmologically formulated.  

Setting the expansion rate to $c$ in the Lemaitre equation \ref{Hubble_eqn}
with $k=0$,
the energy density during dark energy de-coherence is given by
\be
\rho_{PT} \: = \:
{3 c^2 \over 8 \pi G_N}  \left ( {c  \over R_{PT} } \right )^2 \, - \, \rho_\Lambda.
\label{rhoFL}
\ee
Similarly, the scale acceleration at the time of this transition can
be determined:
\be
{ \ddot{R}_{PT} \over R_{PT} } \: = \:
- c^2 \left ( {1 \over R_{PT} ^2 }+ {\Lambda \over 3}
\right )
\Rightarrow  \ddot{R}_{PT} \cong -{c^2 \over R_{PT}}.
\label{PTaccel}
\ee
Since the FL density at de-coherence is specified in terms of the single
parameter given by the phase transition scale $R_{PT}$, all results
which follow depend at most on this single parameter.

To estimate the time of the transition,if the behavior of the cosmology 
can be described as if it remains
radiation dominated in the standard way down to $t=0$, then the scale
parameter satisfies
\be
R(t) \: = \: R_{PT} \left ( {t \over t_{PT}} \right ) ^{1/2} ,
\ee
which gives a time scale at de-coherence as
\be
t_{PT} \: = \: {R_{PT} \over 2 c} .
\ee
The assumption of radiation dominance during de-coherence corresponds to a thermal
temperature of
\be
g(T_{PT}) \: (k_B T_{PT})^4 \: \cong  \:
{90 \over 8 \pi^3} (M_P c^2)^2 \left ( {\hbar c \over R_{PT}} \right ) ^2 ,
\ee
where $g(T_{PT})$ counts the number of degrees of freedom associated
with particles of mass 
$m c^2 << k_B T_{PT}$, and $M_P=\sqrt{\hbar c/G_N}$ is the Planck mass.  Here
we have used Eq. \ref{rhoFL} and the energy density for relativistic thermal
energy $\rho_{thermal}= g(T) {\pi ^2 \over 30} {(k_B T)^4 \over (\hbar c)^3}$.

\subsection{Spatial Curvature Constraints}
\indent

The energy density during dark energy de-coherence $\rho_{PT}$ can be directly determined
from the Lemaitre equation \ref{Hubble_eqn} to satisfy
\be
H_{PT} ^2 =
\left( {c \over R_{PT}} \right )^2 \: = \: {8 \pi G_N \over 3 c^2} 
\left ( \rho_{PT} + \rho_\Lambda \right ) -
{k c^2 \over R_{PT} ^2}.
\label{decoherence}
\ee
A so called
``open" universe ($k=-1$) is excluded from undergoing this transition, since
the positive dark energy density term $\rho_\Lambda$ already excludes a solution
with $\dot{R} \leq c$.
Likewise, for a ``closed" universe that is initially radiation dominated,
the scale factors corresponding to de-coherence $\dot{R}_{PT}=c$
and maximal expansion $\dot{R}_{max}=0$ can be directly compared.
From the Lemaitre equation
\be
{c^2 \over R_{max}^2} \: = \: { 8 \pi G_N \over 3 c^2} \left [ \rho (R_{max}) + \rho_\Lambda \right ] \: \cong \:
{ 8 \pi G_N \over 3 c^2} \rho_{PT} {R_{PT} ^4 \over R_{max} ^4} \Rightarrow
R_{max}^2 \: \cong \: 2 R_{PT} ^2 .
\ee
Clearly, this closed system never expands much beyond the transition scale.
Quite generally, the constraints of quantum measurability for quantized
energy scales ($\dot{R}_\epsilon \le c$)
\emph{requires} that all cosmologies which develop structure be spatially flat.

The evolution of the cosmology during the period
for which the dark energy density is constant and de-coupled 
from the FL energy density is
expected to be accurately modeled using the FL equations.  There is a period of
deceleration, followed by acceleration towards an approximately De Sitter expansion.
The rate of scale parameter expansion is sub-luminal during a
finite period of this evolution, as shown in Figure \ref{redsrate}.
\onefigure{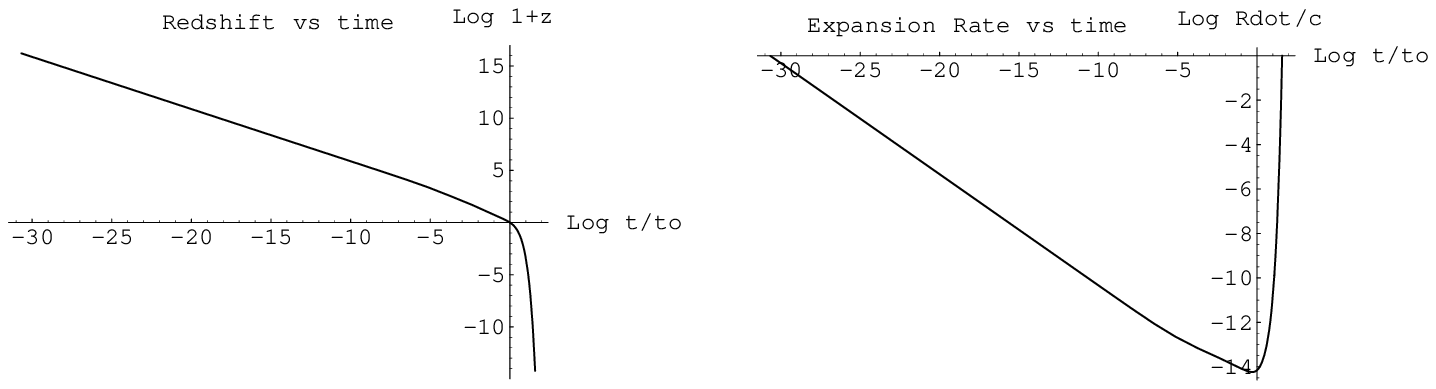}{Graphs of redshift and expansion rate vs time}
The particular value for the scale at de-coherence (which is determined by
the microscopic dynamics of the dark energy during de-coherence)
chosen for the graphs
is given in terms of the measured dark energy density
$\rho_\Lambda = \epsilon^4$.
The present time since the ``beginning" of the expansion
corresponds to the origin on both graphs.
The value of the expansion rate is by assumption equal to the speed of light for
any particular value chosen for $R_{PT}$, as well as when this expansion
scale reaches the De Sitter radius $R_\Lambda$.

\subsection{Estimate of size of source masses for the zero-point energy}
\indent

To estimate the energy scales associated with the zero-point
motions, assume that the dark energy is due to the
uncertainty principle fluctuations of sources with quantized
energy units $M$ (cf. discussion  in Section \ref{intro}
of Casimir effect beginning with Eq. \ref{Casimir}). In general there are
$N$ such sources in a volume specified by $R_{\epsilon} ^3$, and
the quantized energy parameter $\epsilon$. 
On average, each coherent energy unit contributes zero-point
energy of the order
\be
{\epsilon \over N} \sim 
{(\Delta P)^2 \over 2M} \geq {\hbar^2 \over 8M (\Delta X)^2},
\ee
where the uncertainty principle has been used in the form
$(\Delta P)(\Delta X) \geq {\hbar \over 2}$.  Replacing the spatial
uncertainty with the coherence scale $\Delta X \sim R_\epsilon$
relates the energy scale to the zero-point energy $M \sim N \epsilon$.
The cosmological density at the time of the phase transition
is given by the ratio total (non-relativistic) energy to the volume
of coherence
\be
\rho_{PT} =
{NMc^2\over R_{\epsilon}^3} \sim  N^2 {\epsilon^4 \over
(\hbar c)^3} = N^2 \rho_{\Lambda},
\ee
demonstrating that the cosmological density is related to the
number of pairs of coherent energies $N (N-1)/2$ undergoing
zero-point motions.  The FL equations determine the density during the
phase transition from Eq. \ref{rhoFL} 
\be 
\rho_{PT} \cong  {3\over 8 \pi}{(M_{P}c^2 \epsilon^2)^2 \over
(\hbar c)^3} 
\ee 
giving direct estimates of the coherent energy units
involved in the zero-point motions
\be
N \sim {M_{P}c^2\over \epsilon} \quad ; \quad
M \sim M_{P}
\ee
That is, if the phase transition occurs in the radiation-dominated
regime and the extrapolation back to it starts from a cosmological
constant characterized by $\epsilon$, the sources of the vacuum
energy \emph{must} be at the Plank mass scale, each with zero
point energy $\sim {\epsilon^2 \over M_P c^2}$ on average, \emph{independent}
of the specific value of $\epsilon$!  (Note also that the scale
factor at the transition is nearly the same as the
Schwartzschild radius for the mass contained within that radius.) 
This gives a clue as to why the dark energy gets frozen in. 
The coherent mass units undergoing zero-point behaviors
have pairwise gravitational couplings $G_N M^2$ of order unity,
whereas the de-coherent energy density during the FL
expansion will consist of masses with considerably smaller
gravitational couplings.  For brevity, the collective modes of the Planck
mass units will here be referred to as gravons.  The zero-point motions
of those coherent energy units correspond to the vacuum energy
of the gravons.  It is here suggested that after the phase transition,
this vacuum energy
de-coheres from the FL energy density of the cosmology, which
undergoes thermal expansion, and the microscopic coherence scale
changes from $R_\epsilon$ to a much smaller scale given by the
Compton wavelength of the energy scale $m_{UV}$ associated
with the critical density of the transition, as will be discussed in section
\ref{specialrho}. 
It is important to recognize that
gravitational affects are strong due to quantum mechanics,
not energy density, since strong gravitational coherence
occurs at a scale much less than the Planck density.

As indicated in Eq. \ref{thermalscale}, a macroscopic scale $R$ can
be generated in terms of microscopic scales $\Delta x$ and $\lambda_m$. 
In Fig. \ref{critdens}, the intuition is that a critical density of the
cosmology must be reached before any coherent microscopic energy scale
can ``de-cohere" into distinct clusters.  Prior to reaching this density, there
is no available ``space" to de-cohere into.  Since the previous discussion
indicates that many mass units are still within the coherence distance
during this phase transition, one suspects a macroscopic quantum
system during this time.  Any microscopic mass $m$ with the coherence scale defined
in Eq. \ref{thermalscale} will have zero-point energies of the order
${(\Delta P)^2 \over 2 m}  \sim \epsilon$ at the time of the phase transition,
 which will red shift as the cosmology expands.
The specific equation of state depends on the details
of the macroscopic quantum system.  For a gravon gas with
energy $(N_g+ {1 \over 2}) \hbar k_g c=(2 N_g + 1)\epsilon$,
the exponent in Eq. \ref{delE} has the value $b=1$.

\subsection{Holographic considerations}
\indent

The entropy of the system during the phase transition period
will be examined to check that expected holographic bounds are not
exceeded.  
The Fleisher-Susskind\cite{Susskind} entropy limit considers a black hole
as the most dense cosmological object of a given size, limiting the entropy according to
\be
S \: \leq \: S_{_{hole} ^{black}} \: = \:
{k_B c^3 \over \hbar}{A \over 4 G_N},
\ee
where $A$ is a light-like bounding area.
For a radiation dominated cosmology at de-coherence,
the entropy is proportional to the number of quanta, and is related
to the energy density using $\left ( {c \over \dot{R}_\epsilon} \right )^2 \cong
{8 \pi G_N \over 3} \rho_{PT}$ by
\be
{S \over V} \: = \: {4 \over 3} {\rho_{PT} \over T_\epsilon} \Rightarrow
S \: = \: {4 \over \pi G_N} {R_\epsilon \over T_\epsilon}.
\ee
Examining this for the instantaneously
light-like area $\dot{R}_\epsilon=c$ given
by $A=4 \pi R_\epsilon^2$
the ratio of the entropy in a thermal environment to the limiting entropy
during thermalization is given by
\be
{S \over A/4G_N} \: \cong \: {4 \over  \pi^2} \,
\left ( {1 \over R_\epsilon T_\epsilon} \right ) 
\sim 10^{-16}.
\ee
Clearly this result satisfies the FS entropy bound period of the phase
transition.

\setcounter{equation}{0}
\section{Scenarios Prior to Phase Transition
\label{scenarios}}
\indent

Using quantum measurability arguments, it has been shown
that any quantized energy scale $E$ cannot satisfy
uncertainty relations prior to the time that
the cosmological scale associated with that energy
$R_E={\hbar c \over E}$ has a subluminal expansion
rate $\dot{R}_E \leq c$.  The latest quantized energy
scale which manifests is associated with the IR
behaviors of the system.  For times prior to when the
UV mode of the system can satisfy quantum measurability
requirements, the behavior of the system is expected to
be anomolous.  Possible scenarios for earlier states of
the universe which could connect to
the description presented above will be examined in this
section.  
The results involving the amplitude of CMB fluctuations and
relative size of dark energy to FL energy are independent of
the physical scenario prior to the luminal expansion rate of the
scale associated with the dark energy.   Therefore, the ideas discussed
in this paper cannot be used to distinguish between these
scenarios.

\subsection{Inflationary prior state}
\indent

If both general relativity and quantum mechanics can
be reliably used to describe the cosmology prior to the
time of the decoupling of the dark energy, quantum measurability constraints
on the gravitational interactions (which have couplings of order
unity) require a change in the state of the energy density. 
One possible scenario would demand that this energy density
be in the form of vacuum energy with respect to the forms
of matter prevalent in the cosmology shortly after the transition. 

Assuming a transition which conserves energy,
if the energy density of the present cosmology during the
phase transition is set by the inflationary  energy
density, the DeSitter scale of the inflation $\Lambda_i$ is defined
in terms of the dark energy scale of the present cosmology
using
\be
\rho_{\Lambda_i} = {\Lambda_i c^3 \over 8 \pi G_N}=
{3 c^2  \over 8 \pi} {(M_P c^2)^2 \epsilon^2 \over (\hbar c)^3}
\Rightarrow \Lambda_i = {3 \over R_\epsilon ^2}.
\ee
The entropy of the DeSitter horizon associated
with this early inflationary cosmology satisfies
\be
S_\epsilon = {A_\epsilon \over 4 G_N}=
\pi \left( {M_P c^2 \over \epsilon}
\right )^2 k_B,
\ee
and the DeSitter temperature during the inflation is given by
\be
k_B T_\epsilon = {1 \over 2 \pi R_\epsilon}=
{\epsilon \over 2 \pi}.
\ee
The scale of the horizon temperature
during the inflation is comparable to the scale of the dark
energy today.
This is precisely the temperature scale that would drive
thermal fluctuations of the same magnitude as the
zero-point fluctuations that produce the dark energy
during the phase transition to the post-inflationary
cosmology of today.  

Using the FL equation for the acceleration Eq. \ref{PTaccel},
the acceleration just after the phase transition is given by
$\ddot{R}_\epsilon= -\epsilon$.  The inflationary scale
just prior to the phase transition has an acceleration
given by $\ddot{R}_\epsilon = +\epsilon$.  The transition
requires a change in the scale acceleration rate of the
order of the dark energy in each scale region of the
subsequent decelerating cosmology.

\subsection{Bjorken multiple inflations
\label{BjMI}}
\indent

It is of interent to consider a possible scenario first suggested by
J. Bjorken\cite{BJdiscuss,Bj03}. 
Since a luminal scale expansion rate plays a key
role in the present paper, there is no reason not to
expect prior and later periods of decelerations and
inflations in this cosmology, as illustrated in Figure \ref{Bjrate}. 
\onefigure{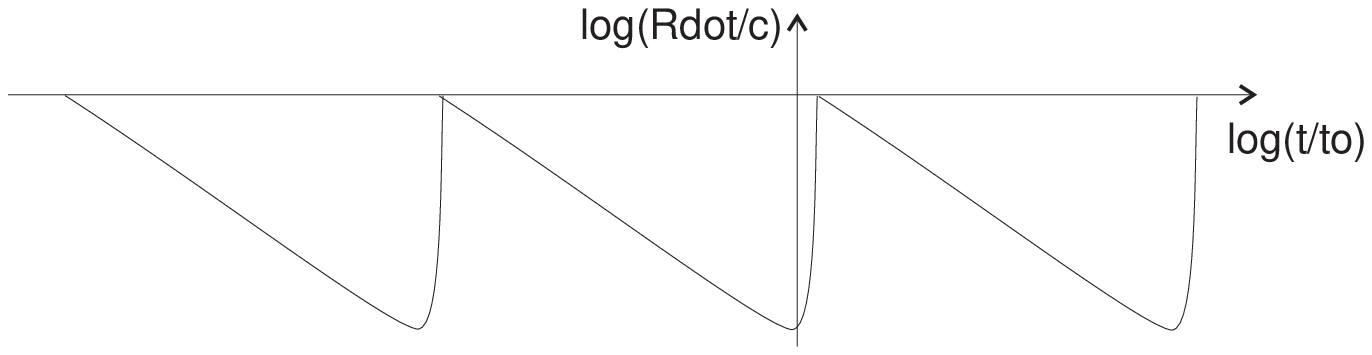}{Multiple inflations limited by subliminal expansion rate}
When relevant scale expansion rate from previous expansion
equals the speed of light, the next deceleration phase begins with the
dark energy density of inflation driving the thermalized matter density
of the next deceleration.  In Bjorken's  scenario,
the landscape of the expansion need not scale uniformly
as shown in Figure \ref{Bjrate}, i.e. valleys need not have
self-similar depths and widths.  The physics of any given deceleration/inflation
cycle is determined by appropriate interaction energy scale of the microscopic
dynamics.  The cycles will maintain self-similarity only if all scales associated
with the phase transitions and broken symmetries scale with the
cosmology.  If this is not the case, the next deceleration will have a
temperature far too cold to produce any of the massive particles
in the spectrum of the standard model.

Assuming that the vacuum density  generated by the cosmological constant
of the prior inflation becomes the FL density in Eq.  \ref{rhoFL}
of new expansion, the phase transition relates the dark energy scales via
\be
\rho_{\Lambda_i} \equiv \epsilon_j ^4 \: = \: 
{3 \over 8 \pi} M_P ^2 \epsilon_{j+1}^2 =\rho_{PT},
\ee
where as before, the dark energy scale satisfies
$
\rho_\Lambda \: = \: {\Lambda \over 8 \pi G_N} 
\: = \: {3 \over 8 \pi G_N R_\Lambda ^2}\: = \: 
{\epsilon ^4 \over (\hbar c)^3 }.
$

Considerable concern has been expressed in the literature
concerning the Poincare recurrences associated with systems
of finite entropy, such as in a DeSitter cosmology\cite{LDJLLS}. 
The entropy associated with a spherically symmetric horizon (while the physics
of that horizon are valid) is given by $S_H=k_B {\pi r_H ^2 \over L_P ^2},$
and the temperature is given by $T_H={\hbar c \over 2 \pi k_B
r_H}$.  A finite entropy, regardless how large, indicates a finite
number of possible configurations for the system, which will
eventually recur into any ``initial" state, eliminating historical
imprints upon that system (along with any information associated
with those imprints).  This return to a prior state has the same
effective outcome as a ``big crunch" in the re-ordering of the system.
To explore whether the present scenario undergoes such recurrences
with any likelihood, examine the entropy change during one of the
transitions between inflation and deceleration:
\be
{S_{j+1} \over S_j} \: = \:
\left ({R_{\Lambda_{j+1}} \over R_{\Lambda_{j}}} \right )^2 \: = \:
{\rho_{\Lambda_j} \over \rho_{\Lambda_{j+1}}} \: = \:
{8 \pi \over 3} {M_P ^2 \over \epsilon_{j+1}^2} \, >> \, 1
\ee
Therefore, since the entropy increases are relatively large,
the time scales of recurrences should be examined to determine
the likelihood of a recurrence prior to the onset of the following
deceleration.  The recurrences are expected
to occur stochastically on time scales given by\cite{LDJLLS, LNANPA04} 
\be
t_{recurrence} \: \cong \: 
t_{reshuffle} \: e^{S_\Lambda \over k_B}, 
\ee 
where $t_{reshuffle}$ is the typical time scale
associated with the microscopic reshuffling of those
configurations that give rise to the finite entropy $S_\Lambda$. 
One expects this reshuffling time (which is directly
related to the mean free time, and is determined by
microscopic dynamics) to be independent of
the number of configurations (except perhaps through
intensive variables like density, etc).
The configurations are counted by the thermodynamic weight
$\mathcal{W}$ in the Boltzmann identification $S=k_B log \mathcal{W}$. This
reshuffling time can be estimated to be a fixed fraction $f_{R}$ of the
causal transit time $\sim {R_\Lambda \over c}$ across the De Sitter
patch (causal region), giving recurrence times of the order 
\be
t_{recurrence} \: \cong \: f_{R} \left ( {R_\Lambda \over c}
\right ) \: e^{\pi \left ( {R_\Lambda \over L_P}  \right ) ^2}
\: \cong \: f_{R} \left ( {R_\Lambda \over c} \right ) \: 
e^{{3 \over 8} \left ( {M_P  c^2\over \epsilon}  \right ) ^4}.
\ee 

An estimate of the time within a given expansion phase can
be estimated using the FL equation for a system with
initial decelerating expansion of order n, followed by inflation
due to the cosmological constant of the epoch:
\be
\left ( {\dot{R} \over R} \right ) ^2 = {8 \pi G_N \over 3}
\left [  \rho_{PT} \left ( {R_{PT} \over R} \right )^n  
+ \rho_\Lambda \right ].
\ee
The duration between times when $\dot{R}=c$ is about
\be
\begin{array}{c}
c \Delta t \cong (1-{2 \over n}) R_\Lambda log{R_\Lambda \over R_{PT}}
+{2 \over n} (R_\Lambda - R_{PT}) + {1 \over 2} R_{PT} \\
\sim R_\Lambda \: log{M_P c^2 \over \epsilon} 
\end{array}
\ee
which is exponentially more rapid than the Poincare recurrence time.
Thus, this scenario essentially eliminates Poincare recurrences,
since the entropy is never bounded, and a recurrence during any
cycle is extremely unlikely.

\subsection{Quantum only cosmology}
\indent

An alternative scenario connects the temporal progression
in classical general relativity to gravitational de-coherence,
implying a single coherent stationary quantum state of density $\rho_{PT}$ prior
to the phase transition.  Since the state is not temporally defined,
the FL equations do not describe the cosmological dynamics
during this ``pre-coherent" period. 
Temporal progression begins only when de-coherent
interrelations break the stationarity of the quantum state,
and is described by the Friedman-LeMaitre equations with
appropriate scale eventually taking the value $R_\epsilon$
as the transition proceeds.  In such a state, there is
no initial temporal singularity, or well defined t=0.  A
quantum fluctuation of a size that produces dark energy $\epsilon$
with scale $R_\epsilon$ gives the same field dynamics as
would an inflationary scenario with DeSitter scale $R_\epsilon$
that matches the phase transition into the decelerating epoch.

Other arguments can be made for expecting a substantial
change in the thermal behavior of the cosmology in early times.
As has been argued, the condition $\dot{R}_E=c$ gives an
IR cutoff for microscopic energy scales that satisfy quantum
measurability constraints.  During earlier times, this cutoff
eventually approaches any microscopic UV cutoff, limiting
the number of available thermal states.  As an illustration,
a bosonic system has a density which satisfies
$\rho=\rho_{IR}+{4 \pi \over (2 \pi \hbar)^3} \int_{k_{IR}} ^{k_{UV}}
k^2 dk {E(k) \over e^{\beta (E-\mu)} -1 }$.  As
the IR and UV cutoffs become comparable,
the thermal behavior is expected to be significantly
different from the usual behavior  with negligible
IR and large UV cutoffs.  An initial state with complete
occupation in a single mode would be expected to be
stationary, and cannot be described with classical
dynamics.

\subsection{Classical relativistic prior state}
\indent

One can examine a scenario for which the violation of the constraints of quantum
measurability results in a classical only dynamics,  
described by the Friedman-Lemaitre
equations of general relativistic cosmology.  Such
a scenario is generally singular at $t=0$.
The question then arises whether any meaning can be given
to a ``radiation dominated" cosmology for a purely classical
system. Since classical general relativity
includes continuous distributions of matter described by an
equation of state, there would seem to be no difficulty in
\emph{defining} such a cosmology in terms of the equation
of state for pressure as one for
which $P= (1/3) \rho$. 
If this scenario is to have ordinary electromagnetic
radiation fields, they would be problematic if used to
define a temperature because of the Rayleigh-Jeans
ultraviolet divergence\cite{Rayleigh}, associated with
the statistics of classical modes instead of quantized
(eg. Planck) modes.  This precludes a statistical description
of this scenario.

\setcounter{equation}{0}
\section{What is Special About $\rho_{PT}$ in \emph{This} Cosmology
\label{specialrho}}
\indent

Up to this point, any value for the critical density $\rho_{PT}$,
and correspondingly for the dark energy scale $\epsilon$, will
generate the observed scale for the amplitude of fluctuations
in the CMB radiation.  The observed scale of dark energy might
be just a random fluctuation.  One might alternatively expect
microscopic physics to fix the particular scale as a quantum
phase transition associated with the UV energy scale for the
gravitational modes, $\rho_{PT} \equiv (m_{UV}c^2)^4/(\hbar c)^3$.  
For instance, if the scale is associated with the
density of thermal bosonic matter, the critical temperature
is related to the (non-relativistic) mass by
$k_B T_{crit}= \left ( {(2 \pi)^2 \over g_m \zeta(3/2)
\Gamma(3/2)}  \right ) ^{2/3}
m_{UV} c^2 \cong  {6.6 \over g_m ^{2/3}} m_{UV} c^2$.
Since this temperature is comparable to the
ambient temperature of thermal standard model matter at
this density, bosonic matter at this density would have a
significant condensate component.   Using
the relation connecting microscopic and macroscopic scale
given previously by $R_\epsilon \lambda_m \sim (\Delta x)^2$,
the zero-point energies of each of the UV energy units 
$m_{UV}$ is of the order of the dark energy 
${(\Delta P)^2 \over 2 m_{UV} } \sim \epsilon$.
Assuming that the cosmological vacuum energy is due
to the zero-point motions of gravitating sources just
at the availability of luminal equilibrations (consistent
with quantum measurability), the dark energy density
can be calculated from the coherence scale using
\be
\rho_\Lambda \: = \: {\epsilon \over R_\epsilon ^3}
\: = \: {\epsilon^4 \over (\hbar c)^3}.
\ee
This means that the cosmological constant gets fixed by
the physical condition of a luminal expansion rate associated
with a microscopic scale (and the associated de-coherence)
being met. 

The currently acccepted values\cite{PDG}
for the cosmological parameters involving
dark energy and matter will be used
for the reverse time extrapolation from the present:
\begin{equation}
h_0 \cong  0.72; \ \ \Omega_{\Lambda} \cong 0.73; \ \ \Omega_M \cong 0.27 .
\end{equation}
Here $h_0$ is the normalized Hubble parameter. Note that this
value implies that the universe currently has the critical energy
density
$\rho_c \simeq 5.5 \times 10^{-4} GeV \ cm^{-3}$.
The values
of these parameters for the present cosmology are given by
$\rho_\Lambda \simeq 4.0 \times 10^{-6} GeV/cm^3, \, 
R_\Lambda \simeq 1.5 \times 10^{28} cm \simeq 1.6 \times 10^{10} ly, \,
\epsilon \simeq 2.35 \times 10^{-12} GeV, \,
R_\epsilon \simeq 8.4 \times 10^{-3} cm, \, 
\rho_{PT} \simeq 1.3 \times 10^{55} Gev/cm^3, \,
m_{UV} \simeq 3133 GeV$.
If the cosmology were to remain a hot radiation dominated
thermal system during the phase transition, with the 
microscopic degrees of freedom $g(T)$ due to the particle spectrum
included, a temperature and redshift for the transition can be
calculated:
\be
\rho_{PT} \sim g(T_{PT}) {\pi^2 \over 30} {(k_B T_{PT})^4 \over
(\hbar c)^3} \Rightarrow
k_B T_{PT} \sim 1300 GeV \: , \: z_{PT} \sim 1 \times 10^{16}.
\ee

\subsection{A connection to microscopic physics}
\indent 
Choosing electro-weak symmetry restoration estimates
of the early 1990's, Ed Jones\cite{Jones90s,Jones97}
predicted a cosmological constant with $\Omega_\Lambda
\simeq 0.6$ before the idea of a non-vanishing small
cosmological constant was fashionable.  Motivated by this
approach, examine symmetry breaking in the early universe \cite{SLACTalk}:
\be
\mathcal{L} = \sqrt{-g} \left [
-{1 \over 2} (D_\mu \Phi_b) g_{\mu \nu} (D_\nu \Phi_b)
+{1 \over 4} m_\Phi ^2 \Phi_b ^2 - {1 \over 8} f^2 \Phi_b ^4
- {1 \over 8} {m_\Phi ^4 \over f^2} 
\right ]  
+ \mathcal{L}_{particle},
\label{Lagrangian}
\ee
where the particle Lagrangian includes the gauge boson field
strength contibution $-F^{\mu \nu} F_{\mu \nu}/ 16 \pi$ 
as well as any vacuum energy subtractions.
In late times, the classical solution where the gauge fields
vanish results in a non-vanishing vacuum expectation value
for one of the field components
\be
|<\Phi_1>|=0 \quad , \quad
|<\Phi_2>|={m_\Phi \over f} = {m_B \over \tilde{e}}.
\label{Higgsvac}
\ee
The constants have been arranged such that as the system
settles into the phase transition, the expected FL energy density is
correctly reproduced with zero contribution of the Higgs field
to the cosmological energy density at late times.
The strategy is to use the energy density due to the
symmetry breaking (Higgs\cite{Higgs}) field prior to the
thermalization of its excitations relative to the
background given in Eq. \ref{Higgsvac} and the vector bosons into the
microscopic particulate states whose remnants
persist today.  The action corresponding to this
Lagrangian
\be
W_{matter} \: \equiv \: \int \mathcal{L} d^4 x
\ee
generate the conserved energy-momentum tensor in
the Einstein equation
\be
T_{\mu \nu} \: = \: -{2 \over \sqrt{-g}}
{\delta W_{matter} \over \delta g_{\mu \nu}}.
\ee

There is every indication that the cosmology will have
extreme spatial homogeneity during the phase
transition, so that for the present, spatial gradients
will be neglected.  For an FRW cosmology, the Jacobian
factor can be calculated using $g=g_{00}g_{xx}g_{yy}g_{zz}$.
The energy density can then be calculated as
\be
T_{00} \: = \: {1 \over 2} ( \dot{\Phi} + \tilde{e} A_0 \Phi )^2 -
{1 \over 4} m_\Phi ^2 \Phi_b ^2 +
{1 \over 8} f^2 \Phi_b ^4 + 
{1 \over 8} {m_\Phi ^4 \over f^2} +
T_{00} ^{particle} .
\label{rhoHiggs}
\ee
When particulate degrees of freedom are negligible,
the general temporal equation of motion for the field
is given by
\be
{1 \over R^3} {d \over dt} \left ( R^3 [\dot{\Phi} + \tilde{e} A_0 \Phi] \right )
- {1 \over 2} (m_\Phi ^2 - f^2 \Phi^2)\Phi = 0,
\ee
which has static solutions when $\Phi$ has vacuum
expectation values of $<\Phi>=0$ and $|<\Phi>|={m_\Phi \over f}$. 
In the earliest stages, when all fields are small, derivatives of the
$\Phi$ field are seen to be of the order $H=\dot{R} / R \sim c/R_\epsilon = \epsilon$,
which is expected to be considerably less than the rates associated with
microscopic transitions involving the vector bosons.
As the field initially evolves, the scalar potential is expected to have
rates which depend on standard model couplings and mass ratios involving
$m_\Phi$ with other particle masses, which should damp rapid
changes in $\Phi$, and smoothes the transition between the ``static" solutions.
The energy density of the system given by $T_{00}$ in Eq. \ref{rhoHiggs}
varies from ${1 \over 8} {m_\Phi ^4 \over f^2}$
for the symmetric solution which has no contribution from the particles
to $ \rho_{PT} + \rho_\Lambda$ when the symmetry is fully broken
and the particles have become manifest.  If the energy density is preserved
during the transition, this gives a relation between the symmetry breaking
parameters and the critical density of the cosmology:
\be
\rho_{PT} \: \cong  \: {1 \over 8} \left ( 
{m_\Phi ^2 \over f}  \right )^2,
\ee
(where $\rho_\Lambda$ has been neglected) which can be solved to give
\be
\begin{array}{c}
m_\Phi = {\tilde{e} \over m_B} \sqrt{8 \rho_{PT}} \\
f \: = \: \left ( {\tilde{e} \over m_B} \right ) ^2
\sqrt{8 \rho_{PT}}.
\end{array}
\ee
Substitution of the form of the expected density during the phase
transition given in Eq. \ref{rhoFL}, the parameters are expected
to satisfy
\be
\begin{array}{c}
m_\Phi =  \sqrt{{3 \over \pi}} {\tilde{e} \over m_B} M_P  \epsilon \\
f \: = \:  \sqrt{{3 \over \pi}} \left ( {\tilde{e} \over m_B} \right ) ^2
M_P  \epsilon.
\end{array}
\ee

If the temporal progression defining the dynamics of
the FL equations only begins after there is de-coherent
energy density present, the initial stationary quantum
state only becomes dynamical once the Higgs field
begins to take a non-vanishing vacuum expectation value. 
However, if both general relativity and quantum properties
hold in the earliest stages, the FL equations have an 
initial inflationary period with the scale parameter satisfying
\be
\begin{array}{l}
\left ( {\dot{R} \over R} \right ) ^2 =
{8 \pi G_N \over 3} \rho_{tot}  \\
\rho_{tot} \cong  \left [
{m_\Phi ^4 \over 8 f^2} +
{1 \over 2} (\dot{\Phi} + \tilde{e} A_0 \Phi)^2 +
{1 \over 8} f^2 \Phi^4 -{1 \over 4} m_\Phi ^2 \Phi ^4
+\rho_{particles}.
\right ]
\end{array}
\ee
The dynamical equation relating the time derivatives of 
the component densities can be obtained from energy
conservation ${T^{0 \mu}} _{;\mu} =0= \dot{\rho}_{tot} +
3{\dot{R} \over R} \: \rho_{tot}$ .  This equation describes
the detailed thermalization of energy into the particulate
states of present day cosmology.

\subsection{Power spectrum considerations}
\indent

A few remarks will be made concerning spatial inhomogeneities
in the cosmology during and after the phase transition.  Since
the fluctuations form after any inflationary periods, no
fine tuning is necessary to prevent amplification of
initial fluctuations\cite{Brandenberger}.
The usual approach\cite{Buonanno}
involves examining metric perturbations $h_{\mu \nu}$ on the classical FRW metric
$(ds^{FRW})^2 = -c^2 dt^2 + R^2(t) \mathbf{dx}^2=R^2(t(\eta)) (-d \eta ^2 +\mathbf{dx}^2) $
in the form $g_{\mu \nu}=g_{\mu \nu}^{FRW}+h_{\mu \nu}$.  Expressing the metric
in terms of the conformal time $\eta$ insures light cones of slope unity in these
coordinates.  The dynamics of the perturbations is most directly calculable using
the form for the curvature tensor $R_{\mu \nu \rho \sigma}={1 \over 2}
(g_{\mu \sigma , \nu \rho}+g_{\nu \rho , \mu \sigma}-g_{\mu \rho , \nu \sigma}-
g_{\nu \sigma , \mu \rho})+{1 \over 2}g_{\lambda \alpha}
(\Gamma_{\nu \rho}^\lambda \Gamma_{\mu \sigma}^\alpha -
\Gamma_{\nu \sigma}^\lambda \Gamma_{\mu \rho}^\alpha)$.
Expanding the small metric perturbations into momentum modes
$h_{\mu \nu}(\eta, \mathbf{x}) \equiv \int d^3 \mathbf{k} e^{i \mathbf{k} \cdot \mathbf{x}}
\tilde{h}_{\mu \nu}(\eta, \mathbf{k})$, the dynamics are given by
\be
{d^2 \tilde{h}_{\mu \nu}(\eta, \mathbf{k}) \over d \eta ^2 }+
2 {R'(\eta) \over R(\eta) } {d \tilde{h}_{\mu \nu}(\eta, \mathbf{k}) \over
d \eta} + k^2 \tilde{h}_{\mu \nu}(\eta, \mathbf{k}) =0
\ee
The dynamics is directly solvable in terms of the reduced perturbations
$\psi_{\mu \nu}(\eta, \mathbf{k}) \equiv R(\eta) \tilde{h}_{\mu \nu}(\eta, \mathbf{k})$,
which satisfies
\be
{d^2 \psi_{\mu \nu}(\eta, \mathbf{k}) \over d \eta ^2 }+
\left [ k^2  -  {R''(\eta) \over R(\eta) } \right ] \psi_{\mu \nu}(\eta, \mathbf{k}) =0.
\ee
This means that short wavelengths that satisfy $k^2>>{R''(\eta) \over R(\eta) }$
will  damp out with the scale factor $ \tilde{h}_{\mu \nu}(\eta, \mathbf{k}) \cong 
 \tilde{h}_{\mu \nu}(\eta_0, \mathbf{k}) e^{|\mathbf{k}| (\eta - \eta_0)} / R(\eta)$,
whereas long wavelength that satisfy $k^2<<{R''(\eta) \over R(\eta) }$ 
will behave as
$\tilde{h}_{\mu \nu}(\eta, \mathbf{k}) \cong  \tilde{h}_{\mu \nu}(\eta_0, \mathbf{k}) +
B_{\mu \nu}(\mathbf{k}) \int_{\eta_0} ^ \eta  {d \zeta \over R^2 (\zeta)}$, which
effectively freezes in those modes at their values at time $\eta_0$.  Since
longer wavelengths satisfy quantum measurability constraints at later times,
the relevant wavelengths should match those of the phase transition.

One expects most of the modes generated by the de-coherence
which form the dark energy to evolve in the same
manner as would those in an inflationary scenario
for times after the horizon crosses the cosmological
scale.  However, UV modes are expected to differ significantly,
due to microscopic de-coherent physics.  Likewise,
the IR modes (either horizon or DeSitter scale)
could differ due to space-like coherence
effects.  Zero-point energies are important in the renormalization
group descriptions of quantum fluids\cite{Berges00}.  Other
authors have found the IR behavior to be important\cite{Gollisch01,Gollisch02}.
More detailed calculations on the power spectrum will be pursued in the future.

\setcounter{equation}{0}
\section{Conclusions and Discussion}
\indent

This paper has presented evidence that current
cosmological observations can be accounted for
by the hypothesis that at early times there was
a phase transition from a macroscopically coherent
state which produced fluctuations consistent with
those observed today. 
Gravitational de-coherence with fluctuations driven by
zero-point motions gives the expected
order for the amplitude of fluctuations in the CMB radiation,
as well as the observed dark energy.
Generally some form of cosmological quantum
coherence (implying space-like correlations and phase coherence)
is expected at densities far from Planck densities. 
Any macroscopic quantum system will manifest such
correlations after it decays or dissolves into luminally
disconnected regions.  These conclusions are
robust in the sense that any theory satisfying these constraints
will produce the same magnitude of CMB fluctuations.

The constraints of quantum measurement (which for instance give the
uncertainty principle) associates with any quantized energy scale
a cosmological scale whose expansion rate is at most luminal.  The
phase transition of interest occurs when the dark energy scale
satisfies this constraint.  Spatial flatness is required for 
observed structure formation if one uses this relevant
microscopic scale in the FL equations.
Using the measured value for the dark energy,
some form of microscopic manifestation of gravitational
physics is expected on the TeV energy scale.

To relate the discussion of the general results to the observed
value of the dark energy density, an argument has been presented to
connect the critical density of the phase transition
to the upper limit of quantized
mass excitations given by the breaking of symmetry using the Higgs
mechanism.  In a ferromagnetic transition, one would expect
excitations with energies comparable to the magnetization
energy to destabilize that magnetization, restoring symmetry. 
Likewise, this critical density could represent the upper limit
on the mass spectrum associated with the Higgs field. 
One should ultimately be able to motivate the form of the Higgs Lagrangian
using density-density fluid dynamics in the early gravitating
epoch.
An exploration of the microscopic physics should give detail to
the power spectrum of the fluctuations generated during the
transition.

Since the results presented depend only on the physics of the
transition, several prior scenarios have been explored which
connect appropriately to the phase transition.  
Quantum coherence reproduces all of the expected results of
an inflationary scenario except for magnetic monopole dilution. 
However, it should be recalled that the very reason for the development
of the theory of special relativity was as an explanation of the
covariance of electromagnetic fields in the absence of external
magnetic sources.  The lack of detection of magnetic monopoles
could simply indicate their non-existence, and has not been taken
as a motivating factor for this paper.  Quite generally,
the phase transition described above ties the top of the
particulate mass spectrum associated with the transition to
the observed value of the cosmological constant.

\begin{center} {\bf Acknowledgment}
\end{center}

We are grateful for long term discussions with E.D.Jones, whose
pioneering work on inflationary microcosmology motivated our interests
in this area.  We also acknowledge several extremely useful discussions
with J.D. Bjorken, who suggested the multi-inflationary scenario discussed
in Section \ref{BjMI}.  Discussions with Stephon Alexander and Michael Peskin have
helped us clarify our arguments concerning quantum constraints on cosmology.
We also extend our gratitude to Walter Lamb and T.W.B. Kibble for discussions
useful to the development of our understanding of cosmological scale.

\end{document}